\documentclass[prl,amsmath,amssymb,showpacs,twocolumn, 	
	      ]{revtex4}
\usepackage{bm}
\usepackage{graphicx}
\usepackage{subfigure}
\usepackage{dcolumn}	

\newcommand{\ri}{\mathrm{i}}

\newcommand{\rd}{\mathrm{d}}
\newcommand{\rJ}{\mathrm{J}}
\newcommand{\bo}{\hat{b}^{\phantom\dag}}
\newcommand{\ba}{\hat{b}^{\dag}}
\newcommand{\Ho}{\hat{H}}
\newcommand{\Qo}{\hat{Q}}
\newcommand{\no}{\hat{n}}
\newcommand{\la}{\langle}
\newcommand{\ra}{\rangle}
\newcommand{\lla}{\langle\!\langle}
\newcommand{\rra}{\rangle\!\rangle}
\newcommand{\be}{\begin{equation}}
\newcommand{\ee}{\end{equation}}
\newcommand{\bes}{\begin{eqnarray}}
\newcommand{\ees}{\end{eqnarray}}

\begin{document}

\title{Avoided level crossing spectroscopy with dressed matter waves}

\author{Andr\'{e} Eckardt$^1$}
\email[email: ]{andre.eckardt@icfo.es}
\author{Martin Holthaus$^2$}
\affiliation{$^1$ICFO-Institut de Ci\`{e}ncies Fot\`{o}niques, 
	E-08860 Castelldefels (Barcelona), Spain}
\affiliation{$^2$Institut f\"ur Physik, Carl von Ossietzky Universit\"at, 
	D-26111 Oldenburg, Germany}
\date{August 14, 2008}

\begin{abstract}
We devise a method for probing resonances of macroscopic matter waves
in shaken optical lattices by monitoring their response to slow parameter
changes, and show that such resonances can be disabled by particular
choices of the driving amplitude. The theoretical analysis of this
scheme reveals far-reaching analogies between dressed atoms and 
time-periodically forced matter waves. 
\end{abstract}



\maketitle

Recently it has been demonstrated experimentally that a macroscopic matter 
wave of ultracold bosonic atoms confined in an optical lattice can be 
controlled in a systematic manner by strong, off-resonant time-periodic 
forcing: Under suitably selected conditions, ``shaking'' the lattice with 
kilohertz frequencies mainly effectuates a modification of the tunneling 
matrix element connecting adjacent lattice sites. In the regime of weak 
interaction, this phenomenon has been inferred from the expansion of a 
Bose-Einstein condensate in a one-dimensional lattice 
geometry~\cite{LignierEtAl07}. A subsequent experiment~\cite{ZenesiniEtAl08} 
utilizes the reduction of the tunneling matrix element to augment the 
relative importance of interparticle repulsion, such that the quantum phase 
transition from a superfluid  to a 
Mott-insulator~\cite{FisherEtAl89,BlochEtAl08} is induced by adiabatically 
varying the amplitude of the driving force~\cite{EckardtEtAl05b}.

These landmark experiments~\cite{LignierEtAl07,ZenesiniEtAl08} clearly 
confirm that there are efficient control mechanisms for ultracold atomic
gases resulting from time-periodic modulation. The situation encountered here 
is akin to the dressed-atom approach: An atom in a laser field becomes 
``dressed'' by that field and changes its behavior~\cite{CohenTannoudjiEtAl98}. Similarly, a many-body matter wave 
becomes dressed in response to time-periodic forcing and acquires properties 
which the unforced, bare matter wave did not have.  

A system of ultracold bosonic atoms in a shaken, sufficiently deep 
one-dimensional optical lattice is described, in the frame of reference
co-moving with the lattice, by the driven Bose-Hubbard model defined by 
the Hamiltonian
$\Ho(t) = \Ho_\text{tun} + \Ho_\text{int} + \Ho_\text{drive}(t)$
\cite{EckardtEtAl05b,CreffieldMonteiro06}. With $\bo_\ell$ and 
$\no_\ell=\ba_\ell\bo_\ell$ denoting the bosonic annihilation and the 
number operator for the Wannier state located at the site labeled by 
$\ell=1,2,\ldots,M$, one has
$\Ho_\text{tun}\equiv -J \sum_{\ell=1}^{M-1}(\ba_\ell\bo_{\ell+1} +
\ba_{\ell+1}\bo_\ell)$, 
where the positive hopping parameter~$J$ implements the kinetics, assumed 
to be exhausted by tunneling between adjacent sites. Moreover,
$\Ho_\text{int}\equiv\frac{U}{2}\sum_{\ell=1}^M\no_\ell(\no_\ell-1)$
with positive interaction parameter~$U$ describes the repulsion of particles 
occupying the same site. Finally,
$\Ho_\text{drive}(t)\equiv K_\omega \cos(\omega t)\sum_{\ell=1}^M\ell\no_\ell$ 
models time-periodic forcing with amplitude~$K_\omega$ and frequency~$\omega$.
With the particle number fixed to $N$, the filling~$n$ is given by the 
ratio~$n\equiv N/M$.

As witnessed by the experiments~\cite{LignierEtAl07,ZenesiniEtAl08}, in a 
time-averaged sense the {\em driven\/} system governed by $\Ho(t)$ behaves 
similar to a system described by the effective, {\em time-independent\/} 
Hamiltonian 
$\Ho_\text{eff}\equiv J_0(K_\omega/\hbar\omega)\Ho_\text{tun}+\Ho_\text{int}$,
which means that the effect of the time-periodic force is captured by
replacing the tunneling matrix element~$J$ by  
$J_\text{eff}\equiv \rJ_0(K_\omega/\hbar\omega)J$, 
with $\rJ_0$ denoting the ordinary Bessel function of order zero. 
This modification of the hopping matrix element is a hallmark of driven 
quantum tunneling~\cite{GrifoniHaenggi98}; it has been cleanly observed 
for single-particle tunneling in strongly driven double-well 
potentials~\cite{KierigEtAl08}. While it becomes exact for a single 
particle on a one-dimensional lattice endowed with nearest-neighbor
coupling~\cite{HolthausHone93}, the dynamics are considerably more 
involved in the many-body case described by the driven Bose-Hubbard model. 
Due to the manifold ways to create excitations in the many-body system,
the $\Ho_\text{eff}$-description is endangered by a multitude of resonances,
and holds approximately only when $\hbar\omega$ is large compared to both 
energy scales which characterize the undriven system,~$U$ and 
$nJ$~\cite{EckardtEtAl05b,EckardtHolthaus07,EckardtHolthaus08}. To further 
explore the newly emerging notion of adiabatic control of driven macroscopic 
matter waves~\cite{ZenesiniEtAl08}, it is now of great importance to study 
such resonances in detail: When do they occur, how strong are they, are they
detrimental to coherent control or can they, perhaps, even be exploited?
These questions mark the scope of the present Letter. By means of numerical 
simulations for small systems, we first outline an experimentally feasible 
detection scheme which allows one to locate major excitation channels in 
parameter space, and to probe their strengths. We also demonstrate that the 
strength of such excitation channels again is subject to coherent control: 
A resonance can be completely quenched by an appropriate choice of the driving 
amplitude. In a second step, we make closer contact between the dressed-atom 
picture and the driven matter waves considered here by studying their 
quasienergy spectrum. In the final third step we explain our findings 
quantitatively by means of perturbation theory for Floquet states.

\begin{figure}[t]\centering
\includegraphics[width = 0.9\linewidth]{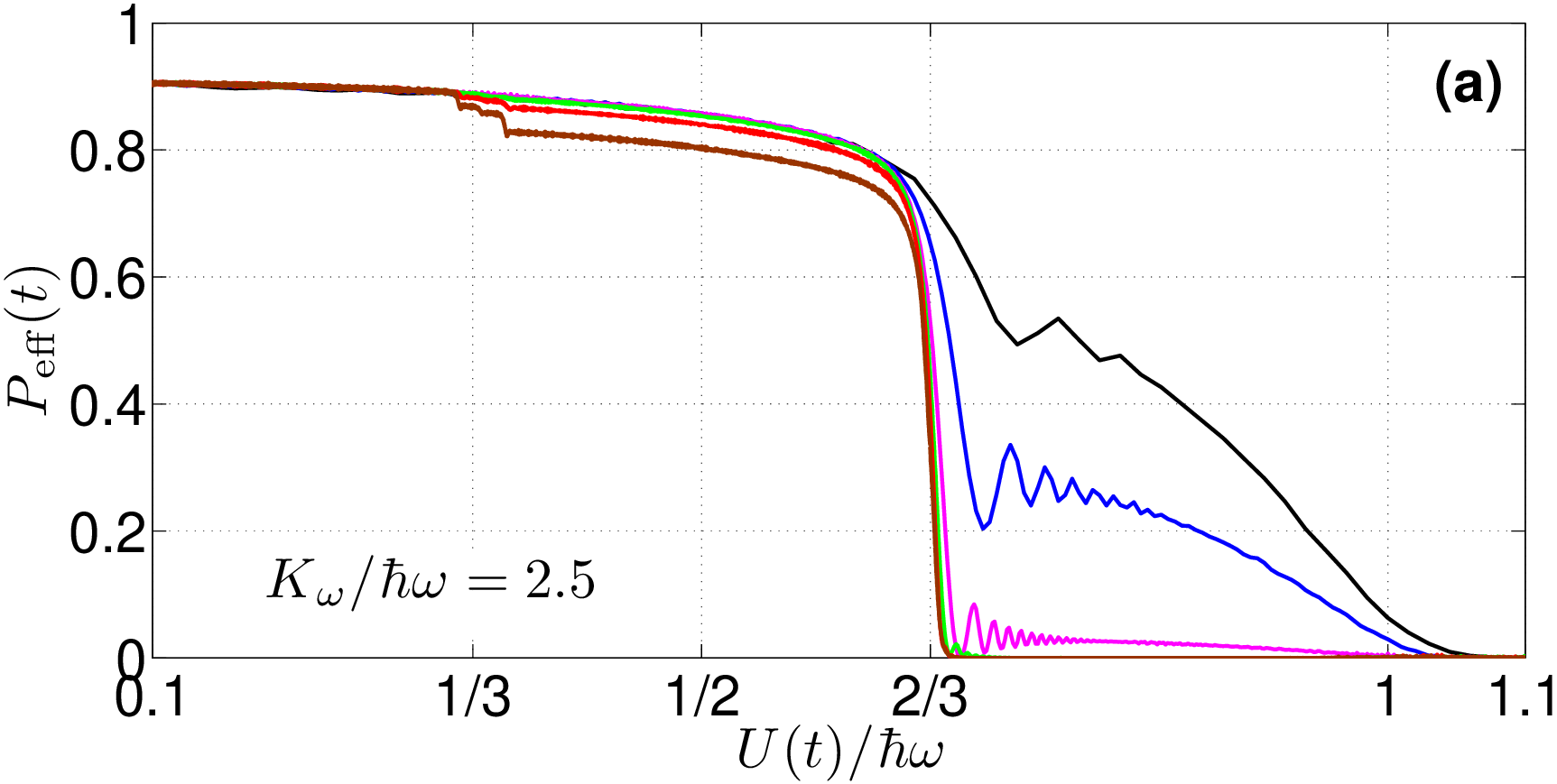}
\includegraphics[width = 0.9\linewidth]{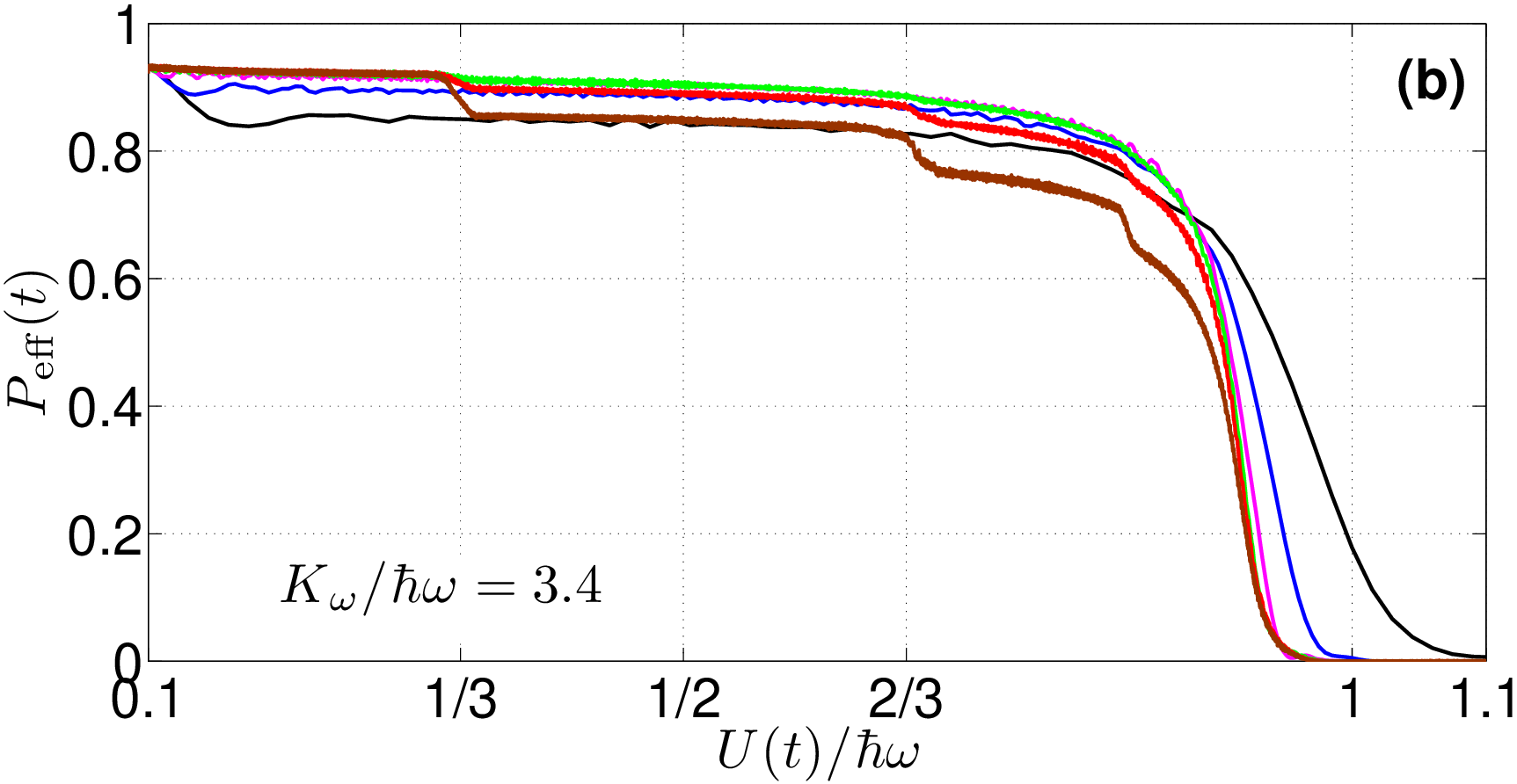}
\caption{\label{fig:F_1} Exact time-evolution of $N = 7$ particles on $ M = 7$ 
	lattice sites. Starting in the ground state at interaction strength 
	$U/J = 0.1$, a drive of frequency $\hbar\omega/J = 20$ has been 
	linearly ramped up within 50 cycles $T = 2\pi/\omega$ to the working 
	amplitude $K_\omega$, before $U$~is increased at various rates 
	$\eta \equiv\dot{U} T/J=$ 0.3 (black), 0.1 (blue), 0.03 (magenta), 
	0.01 (green), 0.003 (red), 0.001 (brown). We plot the squared overlap 
	$P_\text{eff}(t)$ of the instantaneous ground state of $\Ho_\text{eff}$
	with the actual time-evolved state versus $U(t)/\hbar\omega$ at 
	integer~$t/T$. For large $U(t)/\hbar\omega$, $P_\text{eff}$ decreases 
	with decreasing~$\eta$. For $K_\omega/\hbar\omega = 2.5$ there is 
	strong resonant excitation at $U/\hbar\omega = 2/3$~(a). For 
	$K_\omega/\hbar\omega = 3.4$ this resonance is quenched, and another 
	one around $U/\hbar\omega = 1$ becomes active~(b).} 
\end{figure}

Consider the following scenario: A system conforming to the undriven 
Bose-Hubbard model $\Ho_\text{tun} + \Ho_\text{int}$ is prepared in its 
ground state for~$U/J=0.1$. Then a drive $\Ho_\text{drive}(t)$ is switched 
on, with an amplitude increasing linearly in time, and a high frequency  
$\hbar\omega/J=20$. Since this drive is sufficently off-resonant, one expects 
the system to adiabatically follow the ground state of $\Ho_\text{eff}$.
After the working amplitude $K_\omega$ has been reached, it is held constant. 
Then the interaction parameter~$U$ is ramped up at constant rate 
$\eta \equiv\dot{U} T/J$ (with $T = 2\pi/\omega$) into the regime where 
resonances should make themselves felt. In a laboratory experiment this 
can be done, {\it e.g.\/}, by increasing the transversal confinement used 
to create the effective one-dimensional geometry. We have simulated this 
protocol for a small system with $N = M = 7$. In Fig.~\ref{fig:F_1} 
we plot the squared overlap 
$P_\text{eff}(t) = | \la \psi_0^{(\text{eff})} | \psi(t) \ra|^2 $
of the system's true state $|\psi(t)\ra$, obtained by solving the full 
time-dependent Schr\"odinger equation governed by $\Ho(t)$, and the ground 
states $|\psi_0^{(\text{eff})}\ra$ of the corresponding instantaneous 
operators $\Ho_\text{eff}$. Figure~\ref{fig:F_1}~(a) is obtained for 
$K_\omega/\hbar\omega = 2.5$. As expected, $P_\text{eff}$ stays close to unity 
even when~$U$ becomes large, thus validating the $\Ho_\text{eff}$-description, 
until at $U/\hbar\omega \approx 2/3$ it decreases suddenly; the drop is the
more pronounced, the lower the rate~$\eta$. This abrupt decrease signals 
resonant excitation. Experimentally, such resonant excitation can be detected 
by time-of flight absorption imaging. It is indicated by a loss of contrast of 
the sharply peaked structures visible in either the single-particle momentum 
distribution~\cite{BlochEtAl08} if the system is in the superfluid regime 
(which may be reached by a further adiabatic parameter variation), or in the 
two-particle momentum correlations~\cite{AltmanEtAl04,FoellingEtAl05} if the 
system is in the Mott-insulator regime. Interestingly, when choosing the 
particular driving amplitude $K_\omega/\hbar\omega = 3.4$, this excitation 
channel is closed, and another one at $U/\hbar\omega \approx 1$ appears in 
Fig.~\ref{fig:F_1}~(b). This second resonance is stronger than the first
one, since the drop is fully developed already for larger~$\eta$. We conclude: 
(i) $\Ho_\text{eff}$ describes the system up to surprisingly large interaction 
strengths~$U$; (ii) the excitation observed at $U/\hbar\omega \approx 2/3$ 
in Fig.~\ref{fig:F_1}~(a), and at $U/\hbar\omega \approx 1$ in 
Fig.~\ref{fig:F_1}~(b), cannot be ascribed to a deviation from adiabatic 
following on the level of $\Ho_\text{eff}$, since the degree of excitation 
increases with \emph{decreasing} parameter variation rate~$\eta$; 
(iii) a resonance can be disabled by adjusting the driving amplitude. Thus, 
by applying this or a similar protocol, both the locations and the strengths 
of resonant excitation channels can be probed.  

We now shed light on the physics underlying this detection scheme, and 
provide an appropriate theoretical framework. Recall that the dressed-atom
approach deals with atoms interacting with a quantized mode of a radiation 
field. Accordingly, the energy level diagram of the combined system features 
identical copies of groups of levels displaced against each other by the 
photon energy~$\hbar\omega$~\cite{CohenTannoudjiEtAl98}. An analogous picture 
for dressed matter waves driven by a classical time-periodic force is obtained 
by quantum Floquet theory~\cite{Shirley65,Sambe73}: Given the Hamiltonian
$\Ho(t) = \Ho(t + T)$, one defines the quasienergy operator
$\Qo \equiv \Ho(t)- \ri \hbar \partial_t$
which acts in the product space $\mathcal{H}\otimes\mathcal{T}$ made up
from the physical state space $\mathcal{H}$ and the space $\mathcal{T}$
of $T$-periodic functions, and solves the eigenvalue problem
$\Qo|u(t)\rra = \varepsilon|u(t)\rra$. 
Because of the periodic boundary conditions in time, the solutions have
the form 
$|u_{\nu,m}(t)\rra \equiv |u_{\nu,0}(t)\rra\exp(\ri m\omega t)$, 
with $\omega = 2\pi/T$ and $m=0,\pm1,\pm2,\ldots$; the label $\nu$ is
chosen such that $|u_{\nu,0}(t)\rra$ connects to the $\nu$-th energy
eigenstate when the driving force vanishes. Hence, the eigenvalues
$\varepsilon_{\nu,m} \equiv \varepsilon_{\nu,0} + m\hbar\omega$,
called quasienergies, repeat themselves with period $\hbar\omega$ on the
energy axis; each state~$\nu$ placing one copy in each ``Brillouin zone''
of width $\hbar\omega$. Going back to the actual state space~$\mathcal{H}$,
the states   
$|\psi_\nu(t)\ra = |u_{\nu,m}(t)\rra \exp(-\ri\varepsilon_{\nu,m}t/\hbar)$ 
form a complete set of solutions to the time-dependent Schr\"odinger
equation.

\begin{figure}[t]\centering
\includegraphics[width = 1\linewidth]{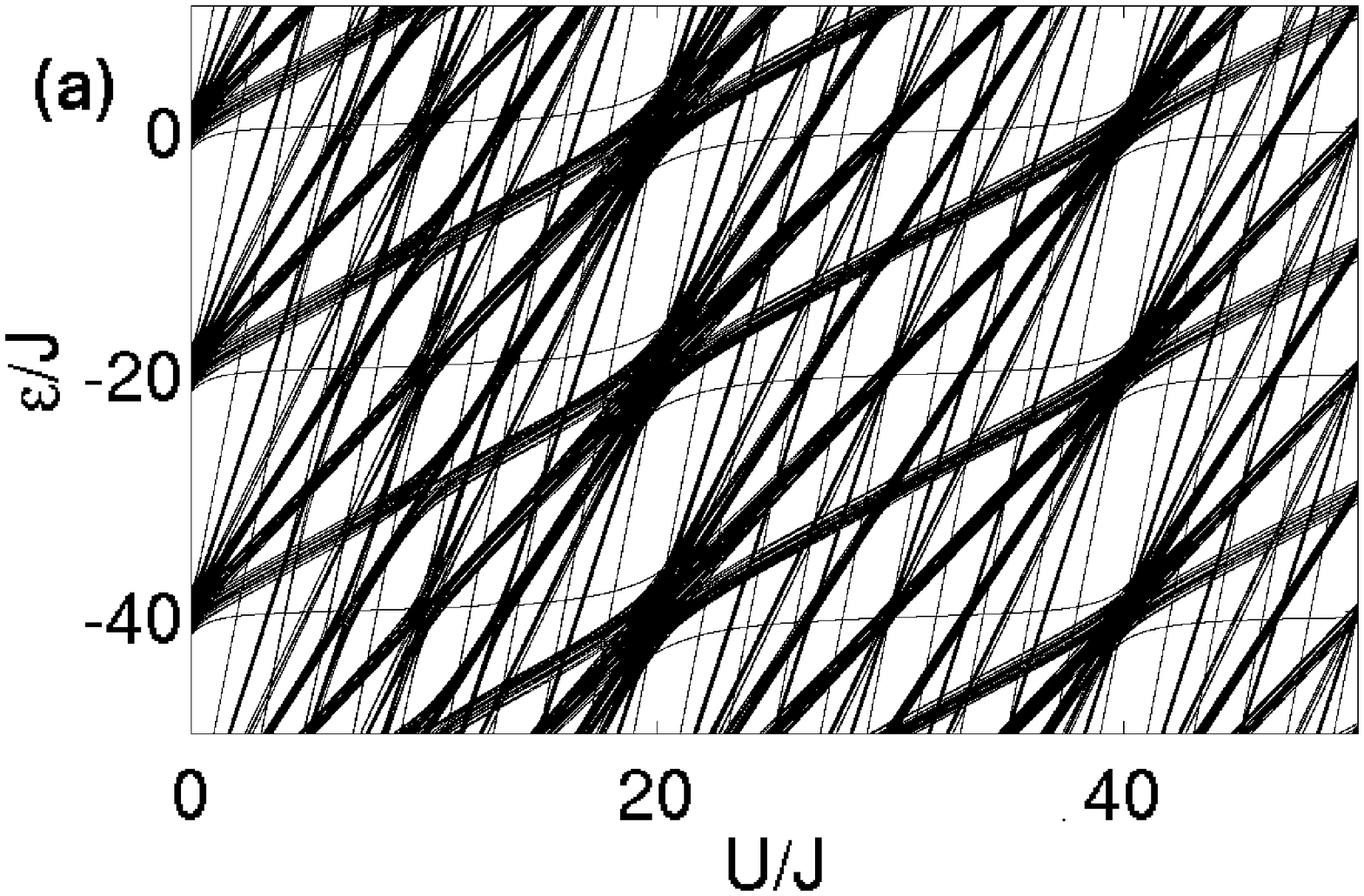}
\includegraphics[width = 1\linewidth]{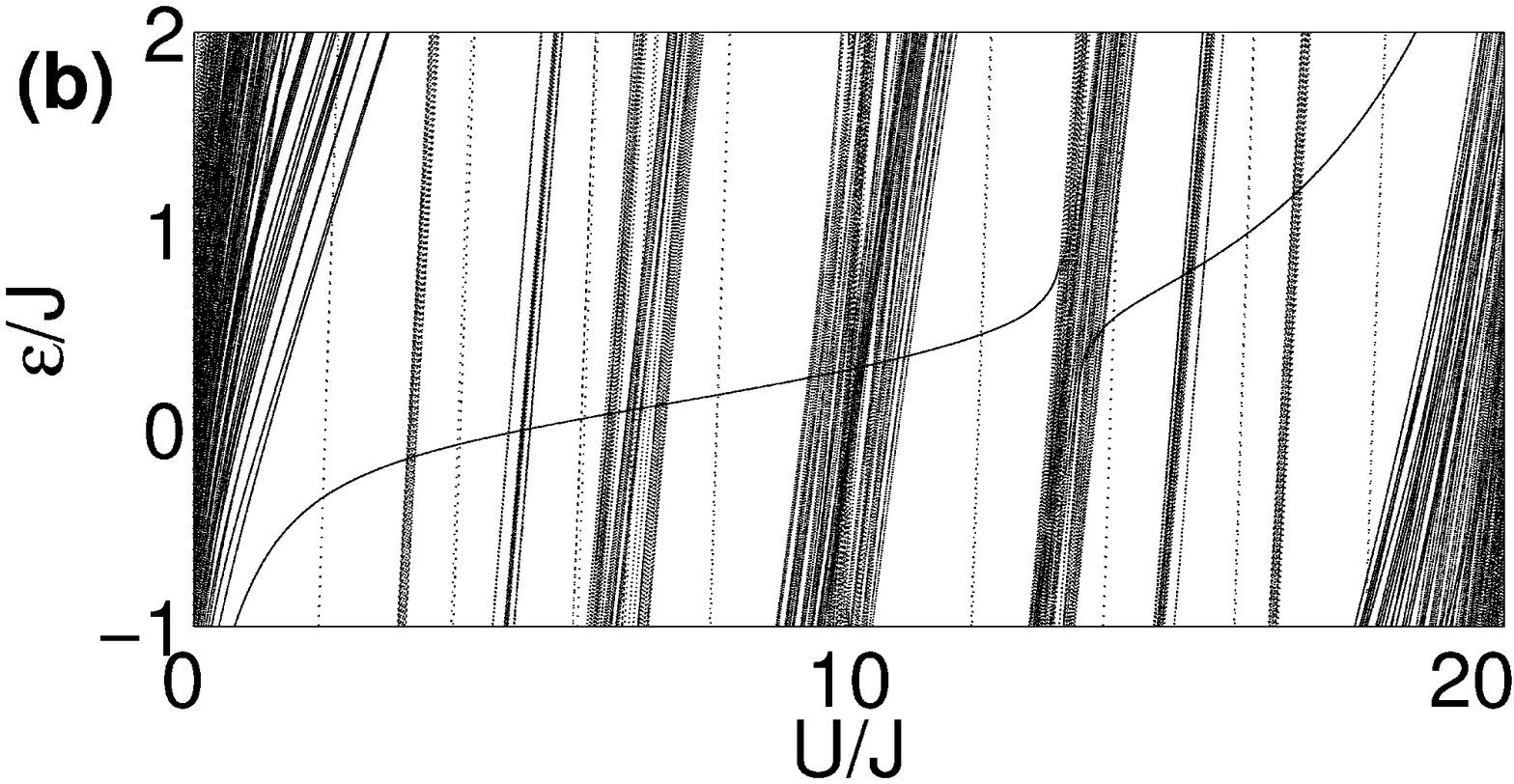}
\caption{\label{fig:F_2} (a) Quasienergy spectrum of a driven 
	Bose-Hubbard system with $N = M = 5$, $\hbar\omega/J = 20$,
        and $K_\omega/\hbar\omega = 2$ versus~$U/J$. Bands with different 
	slopes belong to different types of particle-hole excitations of 
	$\Ho_\text{eff}$. Resonant coupling of such bands results in avoided 
	crossings. The isolated quasienergy level, highlighted in~(b), emerges 
	from the ground state of the undriven system. Clearly visible are the 
	avoided crossings at $U/\hbar\omega \approx 2/3$ ($U/J \approx 13$)
	and $U/\hbar\omega \approx 1$ ($U/J \approx 20$) which have been 
	detected dynamically in Fig.~\ref{fig:F_1}, whereas there are no 
	avoided crossings at $1/3$ and $1/2$.}   
\end{figure}

Fig.~\ref{fig:F_2} shows a part of the quasienergy spectrum belonging to 
a small driven Bose-Hubbard system ($N = M = 5$) with $\hbar\omega/J = 20$ 
and $K_\omega/\hbar\omega = 2$ versus $U/J$. Its basic structure, shown in 
subplot~(a), can be understood as a superposition of copies of the energy 
spectrum of $\Ho_\text{eff}$, shifted against each other by integer multiples 
of $\hbar\omega$. The spectrum of $\Ho_\text{eff}$ possesses bands, made up 
from various types of particle-hole excitations with energies roughly
corresponding to integer multiples of~$U$, clearly identifiable through
their slopes. While in Fig.~\ref{fig:F_2}~(a) quasienergy levels belonging 
to different copies of the $\Ho_\text{eff}$-spectrum hardly ``notice'' each 
other for interaction strengths~$U/J$ much smaller than $\hbar\omega/J = 20$, 
there are pronounced avoided crossings when $U/J$ becomes comparable to 
$\hbar\omega/J$, prominently exemplified by the complex patterns which appear 
when $U/J$ is an integer multiple of $\hbar\omega/J$. Such avoided crossings 
indicate resonances which emerge if eigenstates of $\Ho_\text{eff}$ are 
energetically separated by an integer multiple of $\hbar\omega$; their size 
quantifies the strength of resonant coupling and determines the degree of 
deviation from the $\Ho_\text{eff}$-description. 

Fig.~\ref{fig:F_2}~(b) shows a detail of Fig.~\ref{fig:F_2}~(a), focusing 
on one of the quasienergy copies corresponding to the ground state of 
$\Ho_\text{eff}$. After separating from the bands of excited states 
with increasing $U/J$, thus indicating the superfluid-to-Mott insulator
transition~\cite{ZenesiniEtAl08,EckardtEtAl05b}, this level crosses several
bands associated with different copies of the $\Ho_\text{eff}$-spectrum 
without being notably affected, until it undergoes a wide avoided crossing 
with such a band at $U/J \approx \frac{2}{3}\hbar\omega/J \approx 13$, and
subsequently an even wider one around $U/J \approx \hbar\omega/J = 20$. 
These avoided crossings explain the excitation observed in Fig.~\ref{fig:F_1}: 
The dynamical detection scheme illustrated by that figure relies on the 
adiabatic principle for Floquet states~\cite{EckardtHolthaus08}. With 
increasing~$U$, the state $|\psi(t)\ra$ adjusts itself to the slowly
changing parameter and thus follows the quasienergy level corresponding to
the ground state of $\Ho_\text{eff}$, until it reaches an avoided crossing
too wide to be passed \emph{diabatically}. Then an incomplete
Landau-Zener transition to the anticrossing state excites the system. 
According to Landau-Zener estimates, and in agreemement with the simulations
depicted in Fig.~\ref{fig:F_1}, the excitation probability increases
exponentially with both the width of the anticrossing and decreasing parameter
speed. Thus, the method of detecting resonances in dressed matter waves by 
monitoring their response to slow parameter changes can be regarded as a kind 
of avoided level crossing sprectroscopy. 

Note that in contrast to the regime of linear response, suitable for probing
properties of the undriven system, here we consider the excitation of a 
system which has already been strongly modified by the driving force, in
a manner described by $\Ho_\text{eff}$. Moreover, besides the wide, ``active'' 
avoided quasienergy crossings there also is a host of tiny avoided crossings, 
reflecting the high density of quasienergies in each Brillouin zone, so that 
effectively adiabatic dynamics on the level of $\Ho_\text{eff}$ actually 
includes fully diabatic Landau-Zener tunneling through these narrow 
anticrossings. In an infinitely large system with a truly dense quasienergy 
spectrum, the existence of a well-defined adiabatic limit cannot, thus, be 
expected~\cite{HoneEtAl97}. However, realistic parameter variations take place 
on finite time-scales, in all likelihood making the system ``blind'' against 
such small features of the spectrum.

We now formalize our reasoning. For each admissible set $\{n_\ell\}$
of site-occupation numbers, we employ the usual Fock states 
$|\{n_\ell\}\ra \equiv \prod_\ell(n_\ell!)^{-\frac{1}{2}}
(\ba_\ell)^{n_\ell}|\text{vacuum}\ra$
for constructing an orthonormal basis of Floquet-Fock 
states
$|\{n_\ell\},\widetilde{m}\rra \equiv|\{n_\ell\}\ra
\exp[-\ri\frac{K_\omega}{\hbar\omega} \sin(\omega t)
\sum_\ell\ell n_\ell]\exp(\ri\widetilde{m}\omega t)$
in $\mathcal{H}\otimes\mathcal{T}$, with $\widetilde{m}$ serving as 
``photon'' index for distinguishing different Brillouin zones. 
Invoking the scalar product 
$\lla\cdot|\cdot\rra\equiv\frac{1}{T}\int_0^T\!\rd t\,\la\cdot|\cdot\ra$, 
the quasienergy operator $\Qo\equiv\Qo_0+\Qo_1$ of the driven 
Bose-Hubbard model possesses the matrix elements
\begin{eqnarray*} && 
	\lla \{n_\ell'\} , \widetilde{m}' |\Qo_0 | 
	\{n_\ell\} , \widetilde{m} \rra
\nonumber \\	&&\qquad
	= \delta_{\widetilde{m}',\widetilde{m}} 
	\la\{n_\ell'\} |\big(\widetilde{m} \hbar\omega +\Ho_\text{int} 
	+ j_0 \Ho_\text{tun} \big)|\{n_\ell\}\ra \; , 
\\ &&
	\lla\{n_\ell'\} , \widetilde{m}' |\Qo_1 |
	\{n_\ell\} , \widetilde{m}\rra
\nonumber \\	&&\qquad
	= (1-\delta_{\widetilde{m}',\widetilde{m}})
	j_{s(\widetilde{m}-\widetilde{m}')}
	\la\{n_\ell'\} | \Ho_\text{tun} | \{n_\ell\} \ra\; ,
\end{eqnarray*}
where $j_\nu \equiv \rJ_\nu(K_\omega/\hbar\omega)$ indicates the Bessel
function of order $\nu$ evaluated at $K_\omega/\hbar\omega$, and
$s\equiv \sum_\ell \ell(n_\ell-n_\ell')$, giving $s=+1$ ($s=-1$) 
if $\Ho_\text{tun}$ tranfers one particle by one site to the left 
(right)~\cite{EckardtHolthaus07,EckardtHolthaus08}. This splitting of
the quasienergy operator is performed such that $\Qo_0$ acts within each
subspace with fixed ``photon'' number $\widetilde{m}$ in a manner
conforming to $\Ho_\text{eff}$; whereas $\Qo_1$ describes the coupling 
between these subspaces.

\begin{figure}[t]\centering
\includegraphics[width = 1\linewidth]{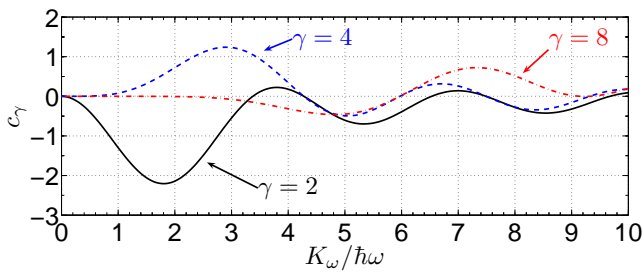}
\caption{\label{fig:F_3} Coupling strength $c_\gamma$ of simultaneous 
	resonant excitation of two particle-hole pairs with the two 
	extra particles located at the same site, evaluated at 
	$U = \gamma\hbar\omega/3$.}
\end{figure}

Let us assume that $U$ is comparable to $\hbar\omega$ while 
$\hbar\omega \gg nJ$, and treat $\Qo_1$ by perturbation theory. 
For $U\gg n|J_\text{eff}|$ the ground state of $\Ho_\text{eff}$ is 
approximately given by the extreme Mott-insulator state 
$|\text{MI}\ra\equiv|\{n_\ell=n\}\ra$ with $n$ particles localized at each 
site. Excited states differ from $|\text{MI}\ra$ by particle-hole 
excitations of energy~$U$; these excitations form bands with widths on the
order of $\sim n|J_\text{eff}|$ due to tunneling of the particles and
holes ``on top'' of $|\text{MI}\ra$. Thus, near $U = \alpha\hbar\omega$ 
with integer $\alpha=1,2,\ldots$ the drive is resonant with respect to
the creation of a single particle-hole pair; eigenstates of $\Qo_0$ differing
from $|\text{MI}\ra$ by one particle-hole pair and $\alpha$ ``photons'' are 
degenerate with $|\text{MI}\ra$ and couple directly ({\it i.e.\/}, in first 
order) via $\Qo_1$ by matrix elements of size  
$-\sqrt{n(n+1)}Jj_{s\Delta\widetilde{m}}$. This coupling leads to the large 
avoided band/level crossings visble in Fig.~\ref{fig:F_2} at $U/J$ close
to 20 and 40.

In second order, the simultaneous creation of two particle-hole pairs via 
(quasi-)energetically distant intermediate states is taken into account.
Intriguingly, second-order coupling between states differing from 
$|\text{MI}\ra$ by $\beta$ ``photons'' and two separate particle-hole 
excitations of total energy $2U$, expected near $U = \beta\hbar\omega/2$ with 
$\beta = 1, 3, 5$ (omitting first-order resonances), vanishes completely 
due to destructive interference between paths involving different intermediate 
states. This explains why there is {\em no\/} avoided crossing at
$U/J \approx 10$ in Fig.~\ref{fig:F_2}. However, there are non-vanishing 
second-order processes creating two \emph{overlapping} particle-hole pairs, 
having two particles or holes sitting at the same site. Assuming unit filling 
$n = 1$, the only possibility is to place both particles at the same site, 
costing the excitation energy~$3U$. For such excitations near
$U = \gamma\hbar\omega/3$ with $\gamma = 1,2,4,5,\ldots$, we find coupling
constants $c_\gamma J^2 n\sqrt{(n+1)(n+2)}/\hbar\omega$ with strengths
$c_\gamma \equiv \frac{1}{2} \sum_{\widetilde{m}' = -\infty}^\infty
(j_{(\gamma+\widetilde{m}')}j_{\widetilde{m}'} + 
j_{-(\gamma+\widetilde{m}')}j_{-\widetilde{m}'})\times
[({2U/\hbar\omega-\gamma-\widetilde{m}'})^{-1}
-({U/\hbar\omega+\widetilde{m}'})^{-1}]$
which vanish for odd~$\gamma$. The plot of $c_\gamma$ depicted in 
Fig.~\ref{fig:F_3} testifies that these strengths depend in an oscillating
manner on the driving amplitude. In particular, it is possible to adjust
that amplitude such that the resonant coupling strength vanishes. For
instance, the zero of $c_2$ at $K_\omega/\hbar\omega \approx 3.4$ is 
the reason for the resonance quenching illustrated in Fig.~\ref{fig:F_1}. 
In $\nu$th~order, coupling matrix elements generally are 
$\sim nJ (nJ/\hbar\omega)^{\nu-1}$; however, we have hardly noticed 
third-order effects in our numerical simulations. Thus, degenerate-state 
perturbation theory in $\mathcal{H}\otimes\mathcal{T}$ systematically 
uncovers the hierarchy of resonances which, in a system with slowly 
changing parameters, become observable order by order with decreasing
parameter speed.  

To conclude, we have outlined a scheme for probing resonances which endanger 
the adiabatic control of macroscopic matter waves achievable through 
time-periodic forcing~\cite{ZenesiniEtAl08}. The theoretical analysis of 
this scheme reveals far-reaching conceptual similarities between dressed atoms 
and dressed matter waves in shaken optical lattices, thus opening up wide new 
grounds between quantum optics and matter-wave physics.

We thank O.~Morsch for many discussions of the 
experiments~\cite{LignierEtAl07,ZenesiniEtAl08}.
A.E.\ is grateful to M. Lewenstein for kind hospitality at ICFO-Institut de 
Ci\`{e}ncies Fot\`{o}niques, and acknowledges a Feodor Lynen research grant 
from the Alexander von Humboldt foundation.

\bibliography{ALS.bib}
\end{document}